\documentstyle[a4,11pt]{article}
\begin{document}

\setlength{\baselineskip}{26pt}

\noindent
{\Large \bf Some observations about the MOLSCAT}

\vspace{4mm}

\setlength{\baselineskip}{18pt}

\noindent
{\Large M.K. Sharma$^a$, Monika Sharma$^a$, and  Suresh 
Chandra$^b$\footnote{corresponding author}}

\noindent
$^a$School of Studies in Physics, Jiwaji University, Gwalior 474 011 (M.P.),
India

\noindent
$^b$Physics Department, Lovely Professional University, Phagwara 144411
(Punjab), India

\noindent
Emails:mohitkumarsharma32@yahoo.in; monika3273@yahoo.in; 
suresh492000@yahoo.co.in

\vspace{6mm}

\noindent
{\bf Abstract.}
For calculation of cross sections for collisional transitions between rotational
levels in a molecule, a computer code, MOLSCAT has been developed by Hutson \& 
Green (1994). For the transitions between rotational levels in H$_2$CS due to
collisions with He atom, we have calculated cross sections under the CS
approximation. In the MOLSCAT, there is provision to input more than one values
of total energies. Here, for example, we are interested in the cross sections
for total energy 11 cm$^{-1}$. The calculations have been done for the single
energy 11 cm$^{-1}$ and for eight combinations, having energies (11, 12), (12,
11), (10, 11), (11, 10), (11, 12, 13), (9, 10, 11), (10, 11, 12), (9, 10, 11,
12, 13) cm$^{-1}$. We have found that the cross sections for 11 cm$^{-1}$, in
general, differ from one another in all the 9 calculations. The reason for the
difference in the results appears that the MOLSCAT uses the intermediate data
of calculations for one energy, in the calculations for other energies. Under
such circumstances, the possible suggestion can be to run the MOLSCAT for a
single energy at a time.

\vspace{2mm}


\section{Introduction}

In most of the cosmic objects, spectral lines of molecules are formed under
non-thermal conditions and for analysis of spectrum, the radiative and
collisional transition probabilities for the transitions between rotational
levels of the molecule are essentially required. For the calculation of 
scattering cross sections for collisional transitions, the first requirement is 
to calculate the interaction potential between the molecule of interest and the
collision partner. Being the most abundant in the interstellar medium, the
hydrogen molecule H$_2$ is taken as the colliding partner. Because of two
hydrogen atoms, the H$_2$ has two species, called the ortho (parallel spins) and
 para (anti-parallel spins). Thus, one needs to consider the collisions between
the molecule of interest and para-H$_2$ (in the $J = 0$ state) or ortho-H$_2$
(in the $J = 1$ state).

The molecule of interest may be a linear molecule (diatomic molecule), symmetric
 top molecule or asymmetric top molecule. Large number of molecules found in the
 cosmic objects belong to the category of asymmetric top molecules. Treatment of
 an asymmetric top molecule is rather complicated, as there is no preferential 
direction. The molecule of our interest, the thioformaldehyde H$_2$CS, is 
asymmetric top molecule.

Often, for simplification of calculations, the molecular hydrogen (collision
partner) is taken as structure-less, and has been replaced by the He atom, as
both the H$_2$ and He have two protons and two electrons, and the interaction
depends on the charges. In the present discussion, we have considered the
collision between the H$_2$CS and He.

In section 2, the interaction potential between the H$_2$CS and He has been
calculated. Section 3 has been devoted for the calculations for collisional
cross sections with the help of MOLSCAT. In the last section 4, we have
discussed about the results and the conclusions have been drawn.

\section{Interaction potential}

For calculation of cross sections for collisional transitions between rotational
levels with the help of MOLSCAT, one requires the interaction potential between
the molecule of interest and the collision partner. For calculation of the
interaction potential between H$_2$CS and He, as the first step, the geometry of
the H$_2$CS molecule has been optimized with the help of GAUSSIAN 2003 and the
coordinates of atoms in the H$_2$CS are obtained. The H$_2$CS is a planar
molecule with electric dipole moment along the axis having the lowest moment of
inertia.

Then, we have included the He atom whose positions have been expressed in terms
of the spherical polar coordinates ($R$, $\theta$, $\phi$) with origin
at the center-of-mass of H$_2$CS. For the
interaction between H$_2$CS and He, we have used the Coupled Cluster with Single
 and Double and perturbative Triple CCSD(T) method and cc-pVTZ basis set. In
order to account for the Basis Set Superposition Errors (BSSE) (Sharma {\it et
al.}, 2014a, b; 2015), we have done three sets of calculations:
\begin{description}
\item{} (i) energy $E_1$ of H$_2$CS + He.
\item{} (ii) energy $E_2$ of H$_2$CS while He is present as a ghost atom.
\item{} (iii) energy $E_3$ of He while all the atoms of H$_2$CS are present as
ghost atoms.
\end{description}

\noindent
There are 106 basis functions, 204 primitive gaussians, 119 cartesian basis
functions. The interaction potential $V(R, \theta, \phi)$ between the H$_2$CS
and He is then
\begin{eqnarray}
V(R, \theta, \phi) = E_2(R, \theta, \phi) + E_3(R, \theta, \phi) - 
E_1(R, \theta, \phi) \nonumber
\end{eqnarray}

\noindent
The interaction potential $V(R, \theta, \phi)$ has been calculated for $R$ =
2.25 (0.25) 5.25 \AA, $\theta =  0^{\circ}$ ($15^{\circ}$) $180^{\circ}$ and
$\phi = 0^{\circ}$ ($15^{\circ}$) $90^{\circ}$. The calculated potential has
been fitted in terms of the spherical harmonics with the help of the
expression:
\begin{eqnarray}
V(R,\theta,\phi) = \sum_{lm} \frac{v_{lm}(R)}{(1+\delta_{m0})} \big[Y_{lm}
(\theta,\phi) +(-1)^m Y_{l-m}(\theta,\phi)\big] \nonumber
\end{eqnarray}

\noindent
where the azimuthal quantum number $l$ has been allowed to vary for the integer
values form 0 to 5. For a given value of $l$, the magnetic quantum number $m$
could assume even integer values from 0 to $l$. The values of the expansion
coefficients $v_{lm}(R)$ as a function of $R$ are given in Table 1.

\begin{table*}
 \centering
\caption{H$_2$CS-He interaction potential in cm$^{-1}$.}
\begin{tabular}{@{}crrrrrr@{}}
\hline\hline                 
$R$ (\AA) & \multicolumn{1}{c}{$v_{00}$} & \multicolumn{1}{c}{$v_{10}$} &
\multicolumn{1}{c}{$v_{20}$} & \multicolumn{1}{c}{$v_{22}$} &
\multicolumn{1}{c}{$v_{30}$} & \multicolumn{1}{c}{$v_{32}$} \\ 
\hline                        
2.25 &  52115.61 & -34010.48 & 47297.66 & -6818.03 & -22737.70 &  9602.23 \\
2.50 & 23789.73 & -15582.89 & 21426.39 & -3027.45 & -10256.35 & 4016.18 \\
2.75 & 10490.22 & -6937.04 &  9373.59 & -1364.29 & -4506.79 & 1729.56 \\
3.00 & 4431.47 & -2931.71 & 3928.66 & -598.34 & -1905.54 &  715.10 \\
3.25 &  1787.34 & -1177.20 & 1579.96 & -254.33 & -779.49 &  280.10 \\
3.50 &  681.02 & -449.18 & 609.07 & -105.13 &  -310.85 &     103.56 \\
3.75 &  237.12 & -160.60 & 222.77 & -42.20 &    -121.20 &      35.63 \\
4.00 &  67.18 & -50.97 & 74.92 & -16.17 & -45.83 &  10.70 \\
4.25 &  6.66 &  -11.77 &  21.12 &  -5.66 &     -16.41 &       2.10 \\
4.50 & -11.84 & 0.65 & 3.10 &      -1.58 &      -5.26 &      -0.46 \\
4.75 & -15.19 & 3.48 & -2.03 &  -0.14 &      -1.26 &      -0.89 \\
5.00 & -13.69 &  3.32 & -2.88 &  0.30 &  0.02 &      -0.74 \\
5.25 & -11.08 &  2.55 & -2.53 &    0.34 &       0.35 &      -0.51 \\
\hline                                   
\hline
$R$ (\AA) & \multicolumn{1}{c}{$v_{40}$} & \multicolumn{1}{c}{$v_{42}$} &
\multicolumn{1}{c}{$v_{44}$} & \multicolumn{1}{c}{$v_{50}$} &
\multicolumn{1}{c}{$v_{52}$} & \multicolumn{1}{c}{$v_{54}$} \\  
\hline                        
2.25 & 2142.69 & -19887.50 & 1152.16 &  6949.83 &   12372.54 &   -2420.97 \\
2.50 & 1259.07 & -8181.70 & 413.10 &    2480.66 &    4807.46 &    -851.89 \\
2.75 & 377.35 & -3557.91 & 167.63 &  1134.88 &    2029.72 &    -349.71 \\
3.00 & 56.16 & -1523.69 & 67.94 &  525.16 &    848.60 &    -144.78 \\
3.25 & -22.72 & -635.48 & 26.52 & 234.40 &     345.14 &     -58.10 \\
3.50 & -28.50 & -259.23 &  9.84 &     100.89 &     137.07 &     -22.45 \\
3.75 & -20.23 & -104.16 &  3.43 &      42.73 &      53.74 &      -8.42 \\
4.00 & -12.32 & -41.10 & 1.09 &  18.05 &      20.84 &      -3.02 \\
4.25 &  -6.80 & -15.59 &  0.24 &  7.44 &       7.80 &      -0.97 \\
4.50 &  -3.42 & -5.47 &  -0.03 &  2.85 &       2.71 &      -0.24 \\
4.75 &  -1.60 & -1.65 &  -0.08 &  1.00 &       0.81 &       0.01 \\
5.00 & -0.65 & -0.30 & -0.06 &       0.29 &       0.16 &       0.05 \\
5.25 & -0.26 &  0.09 & -0.05 &       0.06 &      -0.03 &       0.06 \\
\hline
\end{tabular}
\end{table*}

For the present investigation, the accuracy of interaction potential does
not matter. However, an interaction potential is required. This interaction
potential has been used as input in the computer code MOLSCAT. When the 
interaction potential is not appropriate for the MOLSCAT, the MOLSCAT does not
 converge and no output is produced. For example, in the calculations of 
Green (1980) and Palma \& Green (1987), the BSSE were not considered. When the
BSSE are considered, the potential would definitely be different. The MOLSCAT
had given results for that potential and would give for new potential also.

\section{Calculations with MOLSCAT}

The MOLSCAT has provision to do calculations under the Infinite Order Sudden
(IOS) approximation, Coupled States (CS) approximation, and Close Coupling (CC)
approach. For scattering in an asymmetric top molecule, these three
approaches can be invoked by choosing the value of ITYPE as 106, 26 and 6,
respectively, in the input file for the MOLSCAT. In the IOS approximation, the
energies of rotational levels in the molecule are neglected in comparison to the
 energy of the collision partner. Therefore, it is valid for high energies of
collision partner. Consequently, the the scientists prefer to use the CS
approximation which is valid for all energies of collision partner.

Though the CC approximation is better than the CS approximation, but it is too
expensive from the computation point of view. A calculations in the CC
approximation takes many times more computer time as compared to that in the CS
approximation. In a large number of calculations, the CS and CC approximations
have been used. Some of the papers where such calculations have been done are:
Cernicharo  {\it et al.} (2011), Daniel {\it et al.} (2014; 2015), Dubernet
{\it et al.} (2006), Dumouchel {\it et al.} (2010), Faure \& Josselin (2008),  Faure  {\it et al.} (2012),
Flower \& Lique (2015), Gotoum {\it et al.} (2011), Machin \& Roueff (2006,
2007), Pottage {\it et al.} (2004), Rabli \& Flower (2010a, b; 2011), Sarrasin
{\it et al.} (2010), Troscompt {\it et al.} (2009),  Wernli {\it et al.}
(2006, 2007a, b), Wiesenfeld \& Faure (2013), Wiesenfeld {\it et al.} (2011).

In the present work, for example, we are interested in the cross sections for
total energy 11 cm$^{-1}$. The calculations have been done under the CS
approximation (ITYPE = 26) where the basis set with JMAX = 14 is used. In the
MOLSCAT, there is a provision to input more than one values of total energies.
In the input file, NNRG is the number of total energies included in the input
file. The calculations have been done for the single energy 11 cm$^{-1}$ and for
 eight combinations, (11, 12), (12, 11), (10, 11), (11, 10), (11, 12, 13),
(9, 10, 11), (10, 11, 12), and (9, 10, 11, 12, 13) cm$^{-1}$, as given in Table
 2. In Table 2, column 3 gives the number of energies (NNRG) given in the input
file. The energies and their sequence are given in column 4.

The cross sections for different sets of energies are denoted by C1, C2, \ldots,
 C9. In C1, the MOLSCAT is run for the single energy 11 cm$^{-1}$. In  C2 and
C3, and in C4 and C5, the sequence of energies have been reversed. In C6, two
energies are after the 11 cm$^{-1}$ and in C7, two energies are before the 11
cm$^{-1}$. In C8, one energy before and one energy after the 11 cm$^{-1}$ have
been taken. In C9, two energies before and two energies after the 11 cm$^{-1}$
are taken. One may consider other combinations also. We assume that these
combinations are sufficient for our investigation. All the parameters (except
NNRG and ENERGY) in the input file for all the combinations are the same.

\begin{table*}
 \centering
\caption{Parameters}
\begin{tabular}{@{}rccl@{}}
\hline
No. & Combination & NNRG & ENERGY (cm$^{-1}$) \\
\hline
1 & C1 & 1 & 11 \\
2 & C2 & 2 & 11, 12 \\
3 & C3 & 2 & 12, 11 \\
4 & C4 & 2 & 10, 11 \\
5 & C5 & 2 & 11, 10 \\
6 & C6 & 3 & 11, 12, 13 \\
7 & C7 & 3 & 9, 10, 11 \\
8 & C8 & 3 & 10, 11, 12 \\
9 & C9 & 5 & 9, 10, 11, 12, 13 \\
\hline
\end{tabular}
\end{table*}

The cross sections have been calculated with the help of MOLSCAT. In Table 3,
we have given the cross sections for 11 cm$^{-1}$. Up to 11 cm$^{-1}$ in
H$_2CS$, there are four para levels (0$_{00}$, 1$_{01}$, 2$_{02}$, 3$_{03}$) and
 two ortho levels (1$_{11}$, 1$_{10}$). The ortho and para species of H$_2$CS
behave as they are two distinct molecules, as there are no transitions between
them. Thus, there are 12 (excitations + deexcitations) transitions between the
para levels and 2 (excitations + deexcitation) transitions between the ortho
levels. The cross sections show random values.

In order to understand the range of variation in nine combinations, for
each transition, we have chosen the maximum cross-section $C_{max}$ and the
minimum cross-section $C_{min}$, and have calculated the percent variation $P$
of $C_{max}$ relative to $C_{min}$ as
\begin{eqnarray}
P = \frac{C_{max} - C_{min}}{C_{min}} \times 100  \nonumber
\end{eqnarray}

\noindent
For example, for the transition 0$_{00} \rightarrow$ 1$_{01}$, we have $C_{max}$
 = 23.611 and $C_{min}$ = 14.809. The value of $P$ for each transition is given
in the last column of Table 3.

\begin{table*}
\centering
\caption{Cross sections for various transitions in H$_2$CS for total energy 11
 cm$^{-1}$ in 9 combinations and percent variation $P$.}
\begin{tabular}{@{}rrrrrrrrrrr@{}}
\hline
& \multicolumn{9}{c}{Cross sections in \AA$^2$} &  \\
\cline{2-10}
Transition & C1 & C2 & C3 & C4 & C5 & C6 & C7 & C8 & C9 & $P$ \\
\hline
0$_{00} \rightarrow$ 1$_{01}$ &  16.737 &  14.809 &  21.488 &  18.060 &  16.737 &  17.836 &  23.611 &  16.946 &  23.032 &  59.4 \\
0$_{00} \rightarrow$ 2$_{02}$ &   8.280 &   9.175 &   8.999 &   9.326 &   8.280 &  11.042 &   7.636 &   9.253 &   7.666 &  44.6 \\
0$_{00} \rightarrow$ 3$_{03}$ &  19.136 &  19.148 &  19.898 &  19.299 &  19.136 &  20.372 &  19.222 &  19.991 &  20.690 &   8.1 \\
1$_{01} \rightarrow$ 0$_{00}$ &   6.227 &   5.510 &   7.995 &   6.720 &   6.227 &   6.636 &   8.785 &   6.305 &   8.569 &  59.4 \\
0$_{00} \rightarrow$ 2$_{02}$ &   8.280 &   9.175 &   8.999 &   9.326 &   8.280 &  11.042 &   7.636 &   9.253 &   7.666 &  44.6 \\
0$_{00} \rightarrow$ 3$_{03}$ &  19.136 &  19.148 &  19.898 &  19.299 &  19.136 &  20.372 &  19.222 &  19.991 &  20.690 &   8.1 \\
1$_{01} \rightarrow$ 0$_{00}$ &   6.227 &   5.510 &   7.995 &   6.720 &   6.227 &   6.636 &   8.785 &   6.305 &   8.569 &  59.4 \\
1$_{01} \rightarrow$ 2$_{02}$ &  24.205 &  24.112 &  32.670 &  27.365 &  24.205 &  23.498 &  29.672 &  26.995 &  27.149 &  39.0 \\
1$_{01} \rightarrow$ 3$_{03}$ &  12.348 &  11.737 &  11.836 &  11.772 &  12.348 &   9.631 &  12.575 &  11.673 &  13.476 &  39.9 \\
1$_{11} \rightarrow$ 1$_{10}$ &  38.799 &  40.058 &  73.123 &  39.529 &  38.799 &  40.875 &  43.286 &  40.311 &  45.526 &  88.5 \\
1$_{10} \rightarrow$ 1$_{11}$ &  40.759 &  42.082 &  76.818 &  41.526 &  40.759 &  42.940 &  45.473 &  42.347 &  47.826 &  88.5 \\
2$_{02} \rightarrow$ 0$_{00}$ &   2.408 &   2.668 &   2.617 &   2.712 &   2.408 &   3.211 &   2.221 &   2.691 &   2.229 &  44.6 \\
2$_{02} \rightarrow$ 1$_{01}$ &  18.919 &  18.847 &  25.536 &  21.389 &  18.919 &  18.367 &  23.192 &  21.100 &  21.220 &  39.0 \\
2$_{02} \rightarrow$ 3$_{03}$ &  38.157 &  36.981 &  29.993 &  35.978 &  31.286 &  33.465 &  32.927 &  42.974 &  30.286 &  43.3 \\
3$_{03} \rightarrow$ 0$_{00}$ &   7.281 &   7.285 &   7.571 &   7.342 &   7.281 &   7.751 &   7.313 &   7.606 &   7.872 &   8.1 \\
3$_{03} \rightarrow$ 1$_{01}$ &  12.627 &  12.002 &  12.104 &  12.037 &  12.627 &   9.848 &  12.859 &  11.936 &  13.781 &  39.9 \\
3$_{03} \rightarrow$ 2$_{02}$ &  49.919 &  48.381 &  39.238 &  47.069 &  40.931 &  43.781 &  43.077 &  56.222 &  39.623 &  43.3 \\
\hline
\end{tabular}
\end{table*}

\section{Discussion and conclusions}

First, we have to state that we have no comment on the the papers where CS and
CC approximations have been used. The only point to be discussed here is that
we have found different cross sections for the same energy (11 cm$^{-1}$) when
different combinations, which depend on the direction also, having 11 cm$^{-1}$
are considered.

Except one pair (between 3$_{03}$ and 2$_{02}$), for other pairs, the cross
sections for C1 and C5 are equal. Other values of cross sections show random
variation. It is obvious that the value of $P$ is the same for excitation and
deexcitation between a pair of levels. It supports the detailed equilibrium
where the rate coefficients for collisional excitation and deexcitation are
proportional to each other. The value of $P$ for ortho transitions is 88.5. It
is quite high and shows that the cross sections can vary up to almost a factor
of 2. It may be because the separation between the levels is very small.

To some extent, the value of $P$ is found to decrease with the increase of
separation between the levels. However, it is not the case for all the
transitions. In absence of any trend shown by the cross sections, it is
difficult to draw any legitimate conclusions. However, it may be suggested that
one should calculate the cross sections for a single value of energy (NNRG = 1)
with the help of the MOLSCAT. It would avoid the possibility of using
intermediate data for the calculations for one energy into the calculations for
other energy.

One may still ask if doing the calculations with MOLSCAT for a single value of
energy (NNRG = 1) is sufficient. We do not find ourselves qualified to make any
comment about it. Probably some one who knows more details about the
MOLSCAT may answer about it. Since there is no substitute for the MOLSCAT, one
has to depend on the MOLSCAT.

\vspace{5mm}

\noindent
{\bf References}

\begin{description}

\item{} Cernicharo, J., Spielfiedel, A., Balanca, C., Dayou, F., Senent, M.-L., 
Feautier, N., Faure, A., Cressiot-Vincent, L., Wiesenfeld, L. \& Pardo, J.R., 
2011. Collisional excitation of sulfer dioxide in cold molecular clouds. A\&A, 
531, A103 (9pp).  

\item{} Daniel F., Faure A., Wiesenfeld L., Roueff E., Lis D.C. \& Hily-Blant 
P., 2014. Collisional excitation of singly deuterated NH$_2$D by H$_2$. MNRAS, 
444, 2544 - 2554.

\item{} Daniel F., Faure A., Dagdigian P.J., Dubernet M.-L., Lique F. \& Forets 
G.P., 2015. Collisional excitation of water by hydrogen atoms. MNRAS, 446, 2312
- 2316.
\item{} Dubernet M.-L., Daniel F., Grosjean A., Faure A., Valiron P., Wernli M.,
 Wiesenfeld L., Rist C., Noga J. \& Tennyson J., 2006. Influence of a new 
potential energy surface on the rotational (de)excitation of H$_2$O by H$_2$ at low
temperature.  A\&A, 460, 323 - 329.

\item{} Dumouchel, F., Faure, A. \& Lique, F., 2010. The rotational excitation 
HCN and HNC by He: temperature dependence of the collisional rate coefficients.
MNRAS, 406, 2488 - 2492.

\item{} Faure A. \& Josselin E., 2008. Collisional excitation of water in warm
astrophysical media. I. Rate coefficients for rovibrationally excited states.
 A\&A, 492, 257 - 264.

\item{} Faure A., Wiesenfeld L., Scribano Y. \& Ceccarelli C., 2012. Rotational
excitation of mono- and doubly-deuterated water by hydrogen molecules. MNRAS, 
420, 699 - 704.

\item{} Flower D.R. \& Lique F., 2015. Excitation of the hyperfine levels of
 $^{13}$CN and C$^{15}$N in collisions with H$_2$ at low temperatures. MNRAS, 
446, 1750 - 1755.

\item{} Gotoum N., Nkem C., Hammami K., Ahamat Charfadine M., Owono Owono L.C.
\& Jaidane N.-E., 2011. Rotational excitation of aluminium monofluoride (AlF) by
He atom at low temperature. Astr. Space Sci., 332, 209 - 217.  

\item{} Green, S., 1980. Collisional excitation of interstellar molecules:
Water, ApJS, 42, 103 - 141.
 
\item{} Hutson J.M., Green S., MOLSCAT computer code, version 14 (1994)
 distributed by Collaborative Computational Project No. 6 of the Engineering and
 Physical Sciences Research Council (UK).

\item{} Machin L. \& Roueff E., 2006. Collisional excitation of mono
deuterated ammonia NH$_2$D by helium, A\&A, 460, 953 - 958.

\item{} Machin L. \& Roueff E., 2007. Collisional excitation of doubly
deuterated ammonia ND$_2$H by helium, A\&A, 465, 647 - 650.

\item{} Palma, A. \& Gree, S., 1987. Collisional excitation of an asymmetric 
rotor, Silican carbide, ApJ, 316, 830 - 835.

\item{} Pottage J.T., Flower D.R. \& Davcis S.L., 2004. The rotational 
excitation of methanol by para-hydrogen. MNRAS, 352, 39 - 43.

\item{} Rabli D. \& Flower D.R., 2010a. The rotational structure of methanol
and its excitation by helium, MNRAS, 403, 2033 - 2044.

\item{} Rabli D. \& Flower D.R., 2010b. The rotational excitation of methanol 
by molecular hydrogen.  MNRAS, 406, 95 - 101.

\item{} Rabli D. \& Flower D.R., 2011. Rotationally and torsionally inelastic
scattering of methanol on helium. MNRAS, 411, 2093 - 2098. 

\item{} Sarrasin E., Abdallah D.B., Wernli M., Faure A., Cernicharo J. \& Lique 
F., 2010. The rotational excitation of HCN and HNC by He: new insights on the
HCN/HNC abundance ratio in molecular clouds. MNRAS, 404, 518 - 526.

\item{} Sharma M., Sharma M.K., Verma U.P. \& Chandra S., 2014a. Collisional 
rates for rotational transitions in H$_2$CO and their application.
 Adv. Space Res., 54, 252 - 260. 

\item{} Sharma M.K., Sharma M., Verma U.P. \& Chandra S., 2014b.
Collisional excitation of vinylidene (H$_2$CC). Adv. Space Res., 54, 1963 -
1971. 

\item{} Sharma M.K., Sharma M., Verma U.P. \& Chandra S., 2015. Collisional 
excitation of thioformaldehyde and Silylidene. Adv. Space Res., 55, 434 - 439.

\item{} Troscompt N., Faure A., Wiesenfeld L., Ceccarelli C. \& Valiron P., 
2009. Rotational excitation for formaldehyde by hydrogen molecules:
ortho-H$_2$CO at low temperature. A\&A, 493, 687 - 696.

\item{} Wernli M., Valiron P., Faure A., Wiesenfeld L., Jankowski P. \& 
Szalewicz K., 2006. Improved low-temperature rate constants for rotational
excitation of CO by H$_2$.  A\&A, 446, 367 - 372. 

\item{} Wernli M., Wiesenfeld L., Faure A. \& Valiron P., 2007a. Rotational
excitation of HC$_3$N by H$_2$ and He at low temperature. A\&A, 464, 1147 - 
1154.

\item{} Wernli M., Wiesenfeld L., Faure A. \& Valiron P., 2007b. Rotational 
excitation of HC$_3$N by H$_2$ and He at low temperature. A\&A, 475, 391 - 391.

\item{} Wiesenfeld L. \& Faure A., 2013. Rotational quenching of H$_2$CO by 
molecular hydrogen: cross sections, rates and pressure broadening, MNRAS, 432, 
2573 - 2578.

\item{} Wiesenfeld L., Scifoni E., Faure A. \& Roueff E., 2011. Collisional
excitation of doubly deuterated ammonia ND$_2$H by para-H$_2$, MNRAS, 413, 509 -
513.

\end{description}

\end{document}